\documentclass[twocolumn,preprintnumbers,amsmath,amssymb,prl,superscriptaddress]{revtex4}
\usepackage{graphicx}
\usepackage{dcolumn}
\usepackage{bm}
\usepackage{psfrag}

\newcommand{\be}{\begin{equation}}
\newcommand{\ee}{\end{equation}}
\newcommand{\rme}{{\rm e}}
\newcommand{\rmi}{{\rm i}}

\begin{document}
\title{Fano interference and cross-section fluctuations in molecular
photodissociation}
\author{Y. Alhassid}
\affiliation{Center for Theoretical Physics, Sloane Physics Laboratory, Yale University, New Haven, Connecticut 06520}
\author{Yan V. Fyodorov}
\affiliation{Department of Mathematical Sciences, Brunel University, Uxbridge UB83PH, UK}
\author{T. Gorin}
\author{W. Ihra}
\affiliation{Theoretical Quantum Dynamics, University of Freiburg, 79104 Freiburg, Germany}
\author{B. Mehlig}
\affiliation{School of Physics and Engineering Physics,
GU/Chalmers, 41296 G\"{o}teborg, Sweden }

\begin{abstract}
We derive an expression for the total photodissociation cross section of a
molecule incorporating both indirect processes that proceed through
excited resonances, and direct processes. 
We show that this cross section exhibits
generalized Beutler-Fano line shapes in the limit of isolated
resonances.
Assuming that the closed system can be modeled by random matrix theory, we derive the statistical properties of the photodissociation cross section and find that they are significantly affected by the direct processes. 
We identify a unique signature of the direct processes in the
cross-section distribution in the limit of isolated resonances.
\end{abstract}
\maketitle

Spectral correlations of closed quantum systems, whose associated classical dynamics are chaotic, are known to be nearly universal 
and can be modeled by the Gaussian invariant ensembles
of random matrix theory~\cite{Bohigas,haa91,guh98}. When such systems
become open through their coupling to continuum channels, their bound states acquire decay widths and become resonances, but they are still expected to exhibit universal statistics \cite{fyo97}.
Examples are the conductance fluctuations in quantum
dots~\cite{alh00} and the statistics of the indirect molecular photodissociation cross section ~\cite{fyo98,alh98}. 
If the coupling is weak, the corresponding resonances
are isolated and are often characterized by a Lorentzian line shape (in quantum dots it is necessary to assume temperatures that are much smaller than the decay width). The statistics of the resonance 
widths are determined by the corresponding statistics of 
the bound states and energies $E_n$ of the closed system. For a classically chaotic system, these statistics can be derived from random matrix theory. In recent years, such a random matrix approach has
been successfully used to model the statistics of resonances \cite{fyo97} 
and cross-section
fluctuations in the photodissociation of classically chaotic
molecules~\cite{kir00,dob96,dob95b,pes97,pes95,rei96}.
A semiclassical treatment was discussed in Refs.~\onlinecite{aga99}
and \onlinecite{Eckhardt}.

\begin{figure}[tb]
\psfrag{n}{\mbox{}\raisebox{2mm}{$\hspace*{-0.1cm}|n\rangle$}}
\psfrag{gnc}{\mbox{}$\gamma_{nc}$}
\psfrag{g}{\mbox{}\hspace*{0.0cm}$|g\rangle$}
\psfrag{y1}[Bl][Bl][1][90]{\mbox{}\hspace*{-6mm}\raisebox{-2mm}{$\sigma(E)$}}
\psfrag{R}{$R$}
\psfrag{y}{\mbox{}\hspace*{-1mm}$E$}
\psfrag{E1}{\raisebox{-1mm}{$e_0(R)$}}
\psfrag{E2}{$e_1(R)$}
\psfrag{E3}{\raisebox{5mm}{\raisebox{1mm}{\hspace*{-2mm}$e_2(R)$}}}
\psfrag{S}{dissociation}
\psfrag{C}{continuum}
\psfrag{x}{\mbox{}\hspace*{-3cm}dissociation coordinate $R$}
\mbox{}\\[5mm]
\centerline{\includegraphics[width=0.55\columnwidth]{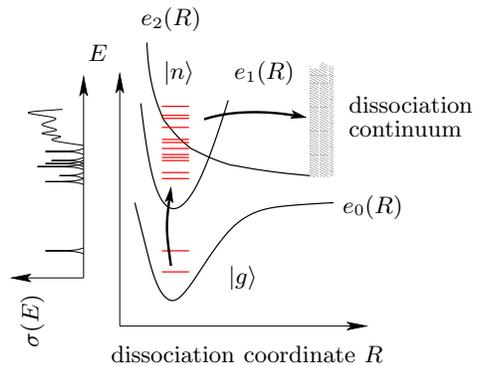}}
\caption{\label{fig:model} Indirect molecular photodissociation.
Shown are three electronic surfaces: the ground-state surface $e_0$  (in which the ground state $|g\rangle$ resides), a binding surface $e_1$ 
and a repulsive surface $e_2$. 
The surfaces $e_1$ and $e_2$ give rise to an effective electronic surface with a barrier near their crossing (not shown), and the bound vibrational states $|n\rangle$ in $e_1$  become  resonances. Indirect photodissociation proceeds through these resonances with a total
cross section $\sigma(E)$ (shown on the left) that is a sum of Lorentzian line shapes.}
\end{figure}

However, as first observed by Beutler~\cite{Beutler} and later 
interpreted by Fano~\cite{fan61}, the
line shape of an individual resonance may differ substantially from 
a Lorentzian: interference between the indirect 
decay via the quasi-bound
state and direct (fast) decay to the continuum gives rise
to a so-called Beutler-Fano line shape. 
For real bound-state wave functions, 
the line shape (versus energy $E$)
is proportional to $(q_n+\varepsilon_n)^2/(1+\varepsilon_n^2)$
where $\varepsilon_n=2(E-E_n)/\Gamma_n$. Here, 
$q_n$ is the Fano parameter characterizing
the line shape, and $\Gamma_n$  is the width of the $n$-th resonance.
Beutler-Fano profiles have been observed in molecular photodissociation~\cite{cot94},
 autoionization~\cite{zak95}, conductance
through quantum wells~\cite{fai97} and quantum dots~\cite{goe00},
STM spectroscopy of surface states~\cite{gad00}, semiconductor
superlattices~\cite{can01}, and Aharonov-Bohm
rings~\cite{kob02}. 

The Fano parameter $q_n$ fluctuates from one resonance to another. Distributions of the Fano parameter have been 
calculated for transmission through a quantum dot~\cite{bro00}
and for the photodissociation of molecules~\cite{ihr02} in the one-channel case (assuming that the corresponding closed system is classically chaotic).

\begin{figure}[bt]
\psfrag{TAG1}{\mbox{}\hspace*{0.0cm}\raisebox{-1mm}{$|c\rangle$}}
\psfrag{TAG2}{\mbox{}\hspace*{-1.mm}\raisebox{-2.5mm}{$\alpha^{\rm in}_n$}}
\psfrag{TAG4}{\mbox{}\hspace*{-0.9cm} \raisebox{-7mm}{$\alpha^{\rm ch}_c$}}
\psfrag{TAG5}{\mbox{}\hspace*{0.0cm}$\gamma_{nc}$}
\psfrag{TAG6}{$|n\rangle$}
\psfrag{TAG7}{$|g\rangle$}
\psfrag{TAG8}{\mbox{}\hspace*{-4mm}$E-E_g$}
\psfrag{A}{}
\psfrag{B}{}
\psfrag{x}{\raisebox{-1mm}{$E$}}
\psfrag{y}{\mbox{}\hspace*{-2mm}\raisebox{2mm}{$\sigma(E)$}}
\mbox{}\\[5mm]
\centerline{\includegraphics[width=0.65\columnwidth]{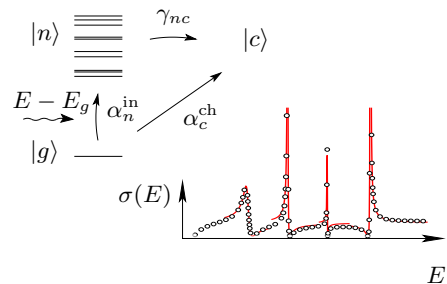}}
\caption{\label{fig:illustration} Top: states
and transition matrix elements in the random matrix model discussed in the text. Bottom: 
the photodissociation cross section $\sigma(E)$ (arb. un.) for one realization of $H_0$ (circles) and for $\beta=1$, $N=128$ and $\Lambda=1$. 
The solid lines describe fitted Beutler-Fano liner shapes.}
\end{figure}

Here we derive an expression for the total photodissociation
cross section for any number of open channels in the presence of
direct decay processes   
(see Figs.~\ref{fig:model} and \ref{fig:illustration}). 
We show that in the limit of isolated resonances their line shapes
have the form of generalized Beutler-Fano profiles
$|q_n+\varepsilon_n|^2/(1+\varepsilon_n^2)$ with a {\em complex} Fano parameter\cite{bro00}
$q_n$.

\begin{figure}[b]
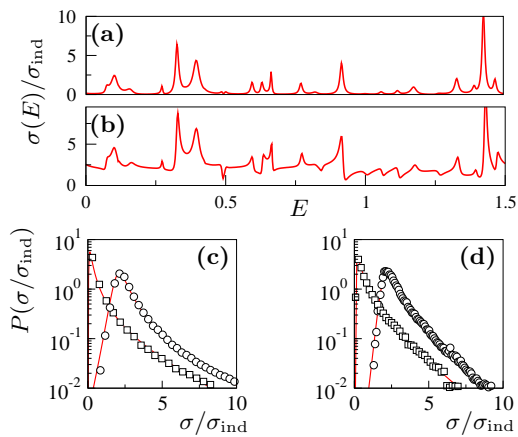

\psfrag{a}{\hspace*{-1mm}\raisebox{-0.5mm}{\bf (a)}}
\psfrag{b}{\hspace*{-1mm}\raisebox{-0.5mm}{\bf (b)}}
\psfrag{XLABEL}{}
\psfrag{YLABEL}{}
\hspace*{-4mm}\centerline{\includegraphics[clip,width=0.79\columnwidth]{sig1.eps}}
\mbox{}\\[0mm]
\psfrag{XLABEL}{\raisebox{5mm}{\hspace*{3mm}$E$}}
\psfrag{YLABEL}{\raisebox{-5mm}{$\hspace*{4mm}\sigma(E)/\sigma_{\rm ind}$}}
\hspace*{-2.5mm}\centerline{\includegraphics[clip,width=0.81\columnwidth]{sig2.eps}}
\psfrag{XLABEL}{\raisebox{1mm}{\hspace*{5mm}$\sigma/\sigma_{\rm ind}$}}
\psfrag{YLABEL}{$P(\sigma/\sigma_{\rm ind})$}
\psfrag{TAG1}{\raisebox{0mm}{$\theta=0$}}
\psfrag{TAG2}{\hspace*{-1mm}$\theta=0.31$}
\psfrag{TAG3}{\mbox{}{\hspace*{7mm}\hspace*{-3.5mm}\bf (c)}}
\mbox{}\vspace*{-4mm}
\centerline{\includegraphics[clip,width=0.35\columnwidth]{h1.eps}
\psfrag{TAG3}{\mbox{}{\hspace*{7mm}\hspace*{-4mm}\bf (d)}}
\psfrag{YLABEL}{}
\hspace*{3mm}
\includegraphics[clip,width=0.35\columnwidth]{h2.eps}}
\mbox{}\\[-6mm]
\caption{\label{fig:hist} 
Cross section distributions.
(a) The cross section vs.\ energy $E$ in the absence of direct coupling ($\theta=0$) and for one random matrix realization of $H_0$. Here $\beta=1, N=64, \Lambda=10$, and $\lambda_c=5\times 10^{-2}$ for all $c$.
(b) Same as in (a) but in the presence of direct processes ($\theta=0.125$). (c) Cross-section distributions at $E=0$
for $\theta=0$ (squares) and $\theta=0.125$
(circles).
(d) as in (c) but for $\beta=2$.
The solid lines in (c) and (d) are the inverse
Fourier transform of  $F_\beta(s)$.
}
\end{figure}

Given that direct photodissociation processes affect the line shapes so dramatically, it is interesting to find out how they affect the statistics of the photodissociation cross-section when the closed system is classically chaotic.
We derive a closed expression 
for the cross-section autocorrelation function (in energy) and find
that it is universal provided that 
the excitation process and the continuum coupling are spatially well-separated. System-specific information enters only in the values
of the direct and indirect channel couplings.
We also calculate the cross-section distribution, and 
find that the
direct decay gives rise to a characteristic maximum 
in the distribution (see Fig.~\ref{fig:hist}) in the regime 
of isolated resonances. This is 
in contrast to the monotonically decreasing behavior of the distribution 
in the same regime, 
in the absence of direct coupling.
In summary, we show that cross-section 
fluctuations are significantly affected by the presence
of direct decay channels, but remain universal.
Our results also apply to atomic autoionization~\cite{zak95,ihr02}.

A molecule can dissociate into several channels $c$ by absorbing a photon.
In the dipole approximation, 
the total photodissociation cross section at energy $E$ is given by
\be
\label{eq:s1}
\sigma(E)= \sigma_0(E) \sum_{c=1}^\Lambda |\langle \Phi_c^{(-)}(E)|\hat \mu|g\rangle|^2
\; ,
\ee
where $\hat \mu = \hat{\bm \mu}\cdot\mathbf{e}$ is the component
of the dipole moment $\hat{\bm \mu}$ of the molecule
along the polarization $\mathbf{e}$ of the absorbed light and
 $\sigma_0(E) \propto (E-E_g)$.
Here $|g\rangle$ is the ground state with energy $E_g$, and
$|\Phi_c^{(-)}(E)\rangle$ ($c=1,\ldots,\Lambda$) is a dissociation
solution at energy $E$ defined by an outgoing wave in
 channel $c$ and incoming waves in all other channels.
We consider a model \cite{fyo97,alh98,fyo98,des99} in which 
the Hilbert space is divided into two parts: an internal
``interacting" region, and an external
``channel" region (cf.\ Fig.~\ref{fig:illustration}). 
The internal region is 
described  by the Hamiltonian $\hat H_0$ represented
by an $N\times N$ matrix ${\bm H}_0$ with eigenstates $|n\rangle$
($n=1,\ldots,N$).  
The external region is spanned by the $\Lambda$ open 
dissociation channels $|c\rangle$. The two regions are 
coupled by an operator $\hat W$ that can be represented by an $N\times \Lambda$ matrix ${\bm W}$ with 
 matrix elements $\langle c|\hat W|n\rangle=\gamma_{nc}$. 
In general, the dipole operator $\hat \mu$ can couple the
ground state to both the internal states $|n\rangle$ 
and the external channels $|c\rangle$.
We define $|\alpha\rangle
 =\hat\mu |g\rangle$, and introduce two vectors ${\bm \alpha^{\rm in}}$ and ${\bm \alpha^{\rm ch}}$. The first has 
  $N$ components
$\alpha^{\rm in}_n \equiv \langle n|\alpha\rangle$, describing the dipole coupling to the internal states, and the second has $\Lambda$ components
$\alpha^{\rm ch}_c \equiv \langle c|\alpha\rangle$, describing the dipole coupling to the continuum channels.

We first show that, in the regime of isolated resonances and 
for ${\bm \alpha^{\rm ch}} \neq 0$, the 
cross section (\ref{eq:s1}) describes a sum over Beutler-Fano resonances.
We write \cite{heller}
$\sum_{c=1}^M |\Phi_c^{(-)}(E)\rangle\langle\Phi_c^{(-)}(E)|=
-\pi^{-1}\mbox{Im}\, \hat G(E+\rmi 0)$ and separate
the channel and internal components of 
the Green function $\hat G(E+\rmi 0)$~\cite{wei}.
We obtain
\begin{eqnarray}
&&\mbox{}\!\!\!\!\!\!\!\!\!\!\!\!\!\!\!
\sigma(E)/\sigma_0(E) = -\frac{1}{\pi}\mbox{Im}
\big[ \langle\alpha|\hat G_{\rm ch}|\alpha\rangle+ \nonumber\\
&&\mbox{}\!\!\!\!\!\!\!\!\!\!\!\!
\langle\alpha|(1\! +\! \hat G_{\rm ch} \hat W^\dagger )
\frac{1}{E\! -\! \hat H_0\! -\! \hat W \hat G_{\rm ch} \hat W^\dagger}
(1\! +\! \hat W \hat G_{\rm ch})|\alpha\rangle\big]\;,
\label{green}\end{eqnarray}
where $\hat G_{\rm ch}$ is the channel Green function. Assuming
unstructured decay continua, Eq.~(\ref{green}) 
simplifies to
\begin{eqnarray}
\label{eq:direct}
\label{eq:cs}
\sigma(E)/\sigma_0(E) &=& 
||{\bm \alpha}^{\rm ch}||^2
\label{unstrc}\\
&&\mbox{}\hspace*{-2.6cm}\!-\!\frac{1}{\pi}\mbox{Im}
\Big[\big({\bm \alpha^{\rm in}}\!\!+\!{\rm i}\pi {\bm W}{\bm \alpha^{\rm ch}}\big )^\dagger
\frac{1}{E\!-\!{\bm H}_{\rm eff}} \big ({\bm \alpha^{\rm in}}\!\!
-\!{\rm i}\pi  {\bm W} {\bm \alpha^{\rm ch}}  \big) \Big]\,.
\nonumber
\end{eqnarray}
Here
${\bm H}_{\rm eff} = {\bm H}_0-{\rm i}\pi {\bm W}{\bm W}^\dagger$ 
is an effective (non-Hermitean) $N\times N$ Hamiltonian in the internal space. 
In the absence of direct photodissociation, ${\bm \alpha^{\rm ch}} = 0$, 
and Eq.~(\ref{unstrc}) reduces to the result of 
Refs.~\onlinecite{alh98} and \onlinecite{fyo98}.

In general, the resolvent in Eq. (\ref{eq:cs}) can be written as 
$(E-{\bm H}_{\rm eff})^{-1} 
= \sum_{n} |R_n\rangle \langle L_n|(E-{\cal E}_n)^{-1}$, 
where $|R_n\rangle$ and $|L_n\rangle$ are 
bi-orthonormal right and left
eigenvectors of ${\bm H}_{\rm eff}$ with complex 
eigenvalues ${\cal E}_n$. However, in the regime of isolated 
resonances we can apply the Breit-Wigner approximation
$|R_n\rangle \approx |L_n\rangle \approx |n\rangle$
and
${\rm Im }\,{\cal E}_n = -\Gamma_n/2 \approx -\pi \sum_c|\gamma_{nc}|^2$.
The cross-section (\ref{eq:cs}) can then be written as a sum over resonances. In the presence of direct photodissociation, the contribution from each of these resonances has a generalized Beutler-Fano line shape  
$|q_n+\varepsilon_n|^2/(1+\varepsilon_n^2)$ with a complex parameter 
$q_n$ whose real part and modulus are given by
\begin{eqnarray}
\label{eq:qn}
{\rm Re}\, q_n &=&
\frac{\raisebox{1mm}{$ {\rm Re}\, [{\alpha^{\rm in}_n}^\ast
\sum_{c=1}^M 
\alpha^{\rm ch}_c \gamma_{nc}$]}}%
{\pi\sum_{c=1}^\Lambda |\alpha^{\rm ch}_c|^2 
\sum_{c=1}^\Lambda|\gamma_{nc}|^2}\,,\\
|q_n|^2 &=&1+\frac{|\alpha^{\rm in}_n|^2-\pi^2\big|\sum_{c=1}^\Lambda\alpha_c^{\rm ch}
\gamma_{nc}\big|^2}{\pi^2\sum_{c=1}^\Lambda |\alpha^{\rm ch}_c|^2
\sum_{c=1}^\Lambda|\gamma_{nc}|^2}\nonumber \,.
\end{eqnarray}
In general, $\mbox{Im}\, q_n \neq 0$, and there is no energy for which the cross section 
vanishes.  However, 
for $\beta=1$ and $\Lambda=1$, Eq.~(\ref{eq:qn}) simplifies 
to Fano's expression 
$\mbox{Re}\,q_n= \alpha^{\rm in}_n/(\pi\alpha^{\rm ch}_c\gamma_{nc})$
and $\mbox{Im}\,q_n=0$
(assuming that all matrix elements are real).

In the following we calculate the statistical properties of the photodissociation cross section (\ref{eq:direct}),  assuming that the dynamics in the closed interaction region are fully chaotic.
The matrix ${\bm H}_0$ is taken to be a $N\times N$
random matrix \cite{Bohigas} from the Gaussian Orthogonal
Ensemble (GOE) or from the Gaussian Unitary
Ensemble (GUE), with distribution
$P({\bm H}_0)\,{\rm d}{\bm H}_0 \propto
\exp[-(\beta N/4) \mbox{Tr} {\bm H}_0^2]\,{\rm d}{\bm H}_0$. Here $\beta = 1$
in the GOE and $\beta=2$ in the GUE.
In the limit of large
$N$, the average eigenvalue distribution (normalized to $1$) is
$\nu(E)=(2\pi)^{-1} \sqrt{4-E^2}$
for $|E|<2$ (and zero otherwise), and 
the corresponding mean eigenvalue spacing 
is $\Delta(E)=1/[N \nu(E)]$. 
In the same limit (i.e., for large $N$),
$\alpha^{\rm in}_n$ and $\gamma_{nc}$ ($c=1,\ldots,\Lambda$) are 
independently distributed Gaussian
random variables, and for each $n$~\cite{alh00}:
\begin{equation}
\label{eq:dist}
P\left(\alpha^{\rm in}_n,{\bm \gamma}_n\right)\!\propto
\exp\left[ 
-\frac{\beta}{2} 
(
{\alpha^{\rm in}_n}^\ast,
{\bm \gamma}_{n}^\dagger 
)\,
{{\bm M}}^{-1}\!\left(\begin{array}{c}
\alpha^{\rm in}_n\\
{\bm \gamma}_{n}
\end{array}\right) 
\right] \;.
\end{equation}
Here ${\bm \gamma}^\dagger_n =(\gamma_{n1}^\ast,\ldots,\gamma_{n\Lambda}^\ast)$ and
\be
\label{eq:tilde-M}
{\bm M} = N^{-1}\, {\bm V}^\dagger {\bm V}\qquad\mbox{with}\quad
{\bm V} = \left({\bm \alpha}^{\rm in},{\bm W}\right)
\ee
is an $(\Lambda+1)\times(\Lambda+1)$ 
matrix. In the following we assume the channel vectors (columns of ${\bm W}$) to be
mutually orthogonal. This can always be achieved by a suitable 
orthogonal (unitary) transformation in channel space.

{\em Average cross section.}
In the center of the band ($E=0$), the 
average cross section is 
$ \langle \sigma\rangle=\sigma_0\, 
\big[\, {||{\bm \alpha}^{\rm in}}||^2/\pi
+\sum_{c=1}^\Lambda |\alpha^{\rm ch}_c|^2/(1\!+\!\lambda_c)\big]
\equiv\sigma_{\rm ind}+\sum_c \sigma_c^{\rm dir}$, where 
$\sigma_{\rm ind}$ and $\sigma_c^{\rm dir}$ are the 
average cross sections in the limiting cases of
purely indirect and purely direct dissociation, respectively.
 Here
$\lambda_1,\ldots,\lambda_\Lambda$ are the $\Lambda$ dimensionless eigenvalues
of the matrix $\pi^2\nu\, {\bm W}^\dagger {\bm W}$.
It is often convenient to characterize the strength of the coupling
to the continuum by transmission coefficients 
$T_c = 4\lambda_c/(1+\lambda_c)^2$.

\begin{figure}[b]
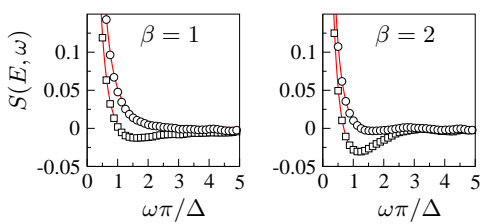

\psfrag{XLABEL}{$\omega\pi/\Delta$}
\psfrag{YLABEL}{$S(E,\omega)$}
\psfrag{TAG1}{$\beta=2$}
\psfrag{TAG2}{$\beta=1$}
\psfrag{L1}{$\theta=0$}
\psfrag{L2}{$\theta=1.26$}
\centerline{\includegraphics[clip,width=0.35\columnwidth]{GOE_largeM.eps}
\psfrag{XLABEL}{$\omega\pi/\Delta$}
\psfrag{YLABEL}{}
\psfrag{L1}{\hspace*{3cm}$\theta=0$}
\includegraphics[clip,width=0.35\columnwidth]{GUE_largeM.eps}}
\caption{\label{fig:largeM} 
Cross-section autocorrelation functions. The symbols are from simulations of random matrices ($N=128$) at the center of the band ($E=0$) and for $\Lambda=10$ and $\lambda_c=10^{-2}$ for all $c$. The  autocorrelation function  without a direct coupling (open squares, $\theta=0$), and with direct coupling ($\theta=1.26$, open circles)
are compared to Eq.~(\ref{eq:Scorr}) (solid lines).
}
\end{figure}

{\em Cross-section autocorrelation function.}
We define a dimensionless cross section 
autocorrelation function~\cite{alh98,fyo98}
 \be
 \label{eq:corr}
 S(E,\omega)\! =\sigma_{\rm ind}^{-2} \big[\big\langle \sigma(E\!-\!\omega/2) 
 \sigma(E\!+\!\omega/2)\big\rangle\!-\!\langle\sigma\rangle^2\big]\,.
 \ee
 In the Breit-Wigner approximation, 
 this correlation is most conveniently calculated in
 the time domain\cite{Levine}. Defining $C(E,t)  = \int_{-\infty}^\infty \!{\rm d}\omega 
 \,{\rm e}^{{\rm i}\omega t}\,S(E,\omega)$, and 
using (\ref{eq:dist}) we find
\be
\label{eq:PEt}
C(E,t) = \frac{1}{4\pi^2}
\left[ A_\beta(t) -B_\beta^2(t) b_{2,\beta}(t)\right]
\ee
where $b_{2,\beta}(t)$ is the two-level form factor~\cite{Bohigas}, and $A_\beta(t)$, $B_\beta(t)$ are functions that depend in general on the matrix ${\bm M}$ in 
Eq. (\ref{eq:tilde-M}) 
and the dipole coupling coefficients to the continuum  $\alpha^{\rm ch}_c$. 

The expression for $A_\beta(t)$ and $B_\beta(t)$ simplify when the dipole ``channel'' is orthogonal to all channel vectors, i.e., ${\bm W}^\dagger {\bm \alpha}^{\rm in} = 0$.  
This is the case when the excitation process and the continuum coupling are spatially well separated.
Using the so-called rescaled Breit-Wigner
approximation \cite{alh98,gorin}, we find 
\begin{eqnarray}
\nonumber
A_1(t) &=& \prod_{c=1}^\Lambda (1+2 T_c\, t)^{-1/2} \Big[
         3+\frac{1}{2}\sum_c \tau_c (1+2 T_c\, t)^{-1}\\[-1mm]
         \label{eq:r1}
         &&\hspace*{5mm}   +\frac{3}{16} \Big(\sum_c \tau_c (1+2 T_c\,
            t)^{-1}\Big)^2\Big]\, \label{rescBW}\\[-2mm]
               B_1(t) &=& \prod_{c=1}^\Lambda (1+T_c\, t)^{-1/2} \Big[ 1-
               \frac{1}{4}\sum_c \tau_c
                  (1+T_c\, t)^{-1}\Big]
                  \nonumber
 \end{eqnarray}
 for $\beta =1$ .
A similar result is obtained for $\beta=2$
\begin{eqnarray}
\nonumber
A_2(t) &=& \prod_{c=1}^\Lambda (1+T_c\, t)^{-1} 
\Big[ 2 + \frac{1}{8}\Big(\sum_c\tau_c\, (1+T_c\, t)^{-1} \Big)^2\Big]\,\\ [-1mm]
\label{eq:r2}\\ [-1mm]
B_2(t) &=& \prod_{c=1}^\Lambda (1+T_c\, t/2)^{-1} \Big[ 1- \frac{1}{4}\sum_c\tau_c\; 
(1+T_c\, t/2)^{-1} \Big] \nonumber\,.
\end{eqnarray}

The results (\ref{eq:r1},\ref{eq:r2}) describe
universal correlations; they depend on the  coefficients $T_c$ and on the parameters  $\tau_c / T_c = \sigma^{\rm dir}_c/
\sigma_{\rm ind}$ which measure the strength of the direct 
photodissociation channels, but they do not depend on the microscopic details of the system (such as the ground state or the nature
of the excitation mechanism).


Two special cases 
 of (\ref{eq:r1}) 
 and (\ref{eq:r2}) 
are of interest. First, in the limit of $\tau_c=0$, the results
of Ref.~\onlinecite{alh98} are recovered, valid in the absence
of direct coupling to the continuum.
Second, consider
the case of $\Lambda$ equivalent open channels 
$T_c = T$, $\sigma_c^{\rm dir} = \sigma^{\rm dir}$. 
In the limit $\Lambda\to\infty$ , $T\to 0$ with $\Lambda T\equiv\kappa$ constant, the
expressions in Eq.~(\ref{rescBW}) simplify to
\begin{equation}
\label{eq:lM1}
A_\beta(t) =  A_\beta \,\rme^{-\kappa\, |t|} \,,\quad
B_\beta(t) = B \,\rme^{-\kappa\, |t|/2}
\end{equation}
with
$A_1 = 3+\theta/2+3\,\theta^2$, $A_2 = 2(1+
\theta/16)$,
$B = (1-\theta/4)$,  
and $\theta = \Lambda \tau = \kappa\, \sigma^{\rm dir}/\sigma_{\rm ind}$. 
We obtain
\begin{eqnarray}
\label{eq:Scorr}
S(E,\omega)&=&\frac{1}{4\pi^2}
\Big[
A_\beta \,f(\omega)\\&&\hspace*{0.5cm}
-B^2\int\!\frac{{\rm d}\omega^\prime}{\pi}\,
f(\omega-\omega^\prime)\, Y_{2,\beta}(\omega^\prime)\Big] \;,   
\nonumber
\end{eqnarray}
where $Y_{2,\beta}(\omega)$ is the two-level cluster
function~\cite{Bohigas},
and $f(\omega)= (\kappa/2)/(\omega^2+\kappa^2/4)$.
Fig.~\ref{fig:largeM} shows Eq.~(\ref{eq:Scorr}) (solid lines)
together with results from random matrix simulations (symbols).
For $\beta=1$ and in the presence
of direct decay, the correlation
function is close to a Lorentzian.

{\em Cross-section distribution.}
The distribution $P\big(\sigma/\sigma_{\rm ind}\big)$ 
is calculated
from its Fourier transform
$F_\beta(s) =\big\langle {\rm e}^{-{\rm i}s\,\sigma/
\sigma_{\rm ind}}\big\rangle$ within the Breit-Wigner approximation.
We have calculated $F_\beta(s)$ for $\Lambda$ equivalent open channels
in the limit of $\Lambda\rightarrow\infty$
with $\Lambda T\equiv \kappa$ kept constant. In this case,
$\Gamma_n \simeq \Gamma=2\Lambda\lambda/N$, and using 
(\ref{eq:dist}) we obtain
\begin{equation}
\label{eq:det}
F_\beta(s) \!= \!
{\rm e}^{{\rm i} s \,\theta/(2\Gamma N)}\,
\Bigg\langle \Bigg(\frac{\det[(E\!-\!{\bm H}_0)^2+\Gamma^2/4]}{\det[(E\!-\!{\bm H}_0)^2+\widetilde\Gamma_{\!\beta}^2/4]}\Bigg)^{\!\frac{\beta}{2}}\Bigg\rangle
\end{equation}
with ${\widetilde\Gamma_{\!\beta}}^2 = \big(\Gamma+4\pi{\rm i}\,s/
(N\beta)\big)
\big(\Gamma-2\pi {\rm i}\,s\, \theta/(N\beta)\big)$. 
Eq.~(\ref{eq:det}) can be evaluated using the results of 
Ref.~\onlinecite{tan}. For $\beta=2$ 
\begin{eqnarray}
F_2(s)&=&
{\rm e}^{{\rm i} s \theta/(2\Gamma N)}
{\rm e}^{-\pi\widetilde\Gamma_{\!2}/\Delta}\nonumber\\
&\times&
\Big[
\mbox{cosh}\big(\frac{\pi\Gamma}{\Delta}\big)
+\frac{1}{2}\,
\mbox{sinh}\big(\frac{\pi\Gamma}{\Delta}\big)
\Big(\frac{\Gamma}{\widetilde\Gamma_{\!2}}+
\frac{\widetilde\Gamma_{\!2}}{\raisebox{-.5mm}{$\Gamma$}}\Big)\Big]\,.
\label{eq:Psig}
\end{eqnarray}
For $\beta=1$ the corresponding result can be expressed
in terms of a four-fold integral~\cite{tan}. Fig.~\ref{fig:hist}
shows the inverse Fourier transform of $F_\beta(s)$ (solid lines) in comparision with random matrix simulations (symbols).  
In the presence of direct coupling, the cross-section 
distribution exhibits a maximum (see Fig.~\ref{fig:hist}b). 
In the limit of isolated resonances, 
this maximum is a clear signature of the direct processes.

In conclusion, we have shown that a direct coupling to the continuum leads to generalized Fano resonances in the total photodissociation cross section, and used random matrix theory to derive the signatures of these direct processes in the cross-section statistics. 
 
 We acknowledge support by a Vice-Chancellor grant
 from Brunel university, by Vetenskapsr\aa{}det, the EU Human Potential Program
under Contract No.\ HPRN-CT-2000-00156, 
and by the U.S. Department of Energy  grant No.\ DE-FG-0291-ER-40608.

\bibliographystyle{prsty}

\vfill

\end{document}